\documentclass{PoS}

\title{Spin dependent gluon distributions and their measurement in heavy quark production processes}

\ShortTitle{Spin dependent gluons}

\author{\speaker{Gary R. Goldstein}
        \\
        Tufts University\\
        Medford, MA, 02155, U.S.A.\\
        E-mail: \email{gary.goldstein@tufts.edu}}

\abstract{Gluon PDFs, GPDs and TMDs play a significant role in an array of scattering processes, including SIDIS, DVCS, exclusive meson electroproduction and p - p scattering. Spin dependent gluon distributions can lead to distinctive features in the angular dependences and asymmetries of the scattering processes. Of particular interest are heavy quark production processes, wherein spin observables of the heavy quarks adumbrate the underlying gluon spin dependences. Top pair production at LHC is a prime example that proceeds primarily via gluon fusion. Decays of polarized top pairs through various channels produce a variety of correlations among the decay products - particles and jets. Combinations of the gluon distributions, either polarized or unpolarized, can be accessed experimentally through angular dependences of decay products, as will be shown, along with predictions from a ``flexible'' spectator model of gluon distributions.}

\FullConference{23rd International Spin Physics Symposium - SPIN2018 -\\
		10-14 September, 2018\\
		Ferrara, Italy}

\usepackage{color,graphicx}% Include figure files
\usepackage{dcolumn}% Align table columns on decimal point
\usepackage{bm}% bold math
\newcounter{comment}

\begin{document}

\newcommand{\be}{\begin{equation}}
\newcommand{\ee}{\end{equation}}  

\section{Introduction}
In the following, I connect three topics that are usually considered separately: gluon distributions; transversity; and top quark pair production. 
There are several versions of the gluon distributions within the nucleon. The most general structures are the Generalized Transverse Momentum Distribution Functions (GTMD's), that depend on the virtual photon+nucleon scattering variables $(x, \xi, \Delta, k_t^2, Q^2)$. Integrations over $k_t^2$ or $\Delta$ reduce these to Generalized Parton Distributions (GPDs) or Transverse Momentum Distributions (TMDs), respectively. Unintegrated models for either of these distributions default back to GTMDs. So the model that will be in the background here, is  a particular extension of a model for GPD's - the Reggeized spectator model referred to as the ``flexible parameterization''  scheme~\cite{GGL}, which has a natural generalization to the gluon and sea quark GPDs~\cite{GLgluons}. The model for gluon GPDs, with parameterization fixed by various constraints, is directly related to the gluon transversity pdf, $h^g_1(x,Q^2)$ and for non-exclusive or semi-inclusive processes (SIDIS), is the TMD $h_1^g(x,\vec{k}_T^2)$.
 
The production and decay of top-antitop pairs in hadron accelerators can provide a measure of the gluon distributions, including the polarization. I will discuss the interesting connection between gluon distributions, with and without polarization, and top-antitop spin correlations. 
The emphasis in this short paper, will be on the top-antitop spin correlations, since these are seldom discussed. Relevant features of the gluon distributions have been discussed in Ref.~\cite{GLgluons} and details will be presented in a forthcoming paper. 

\section{Top-Antitop Spin Correlations}
Before the discovery of the top quark at the Fermilab Tevatron, one proposed method to disentangle the signal for top quark production from the daunting background of multiple hadron events was to concentrate on the spin correlations of the top and antitop decay products. The ``golden events'' were expected to be the dilepton events in which two very energetic opposite sign leptons would signal the weak decays of each top into b-quarks and W's, the latter decaying leptonically.  The actual observations of top quarks by the D0~\cite{D0} and CDF~\cite{CDF} groups did not use the spin correlations. Nevertheless, these correlations provide a test of the QCD mechanism~\cite{spin96}.
The LHC now produces many more top quarks. The higher energy makes quark-antiquark annihilation less important than  gluon fusion. Gluon fusion, involving the merging of two vector particles, gives rise to quite distinct spin correlations among the top decay products. I will present the corresponding spin density matrices and angular correlations. 

\begin{figure*}
\includegraphics[width=8.1cm]{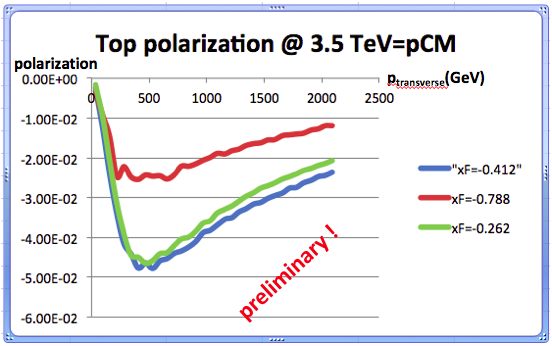}
\hspace{0.05mm}
\includegraphics[width=8.1cm]{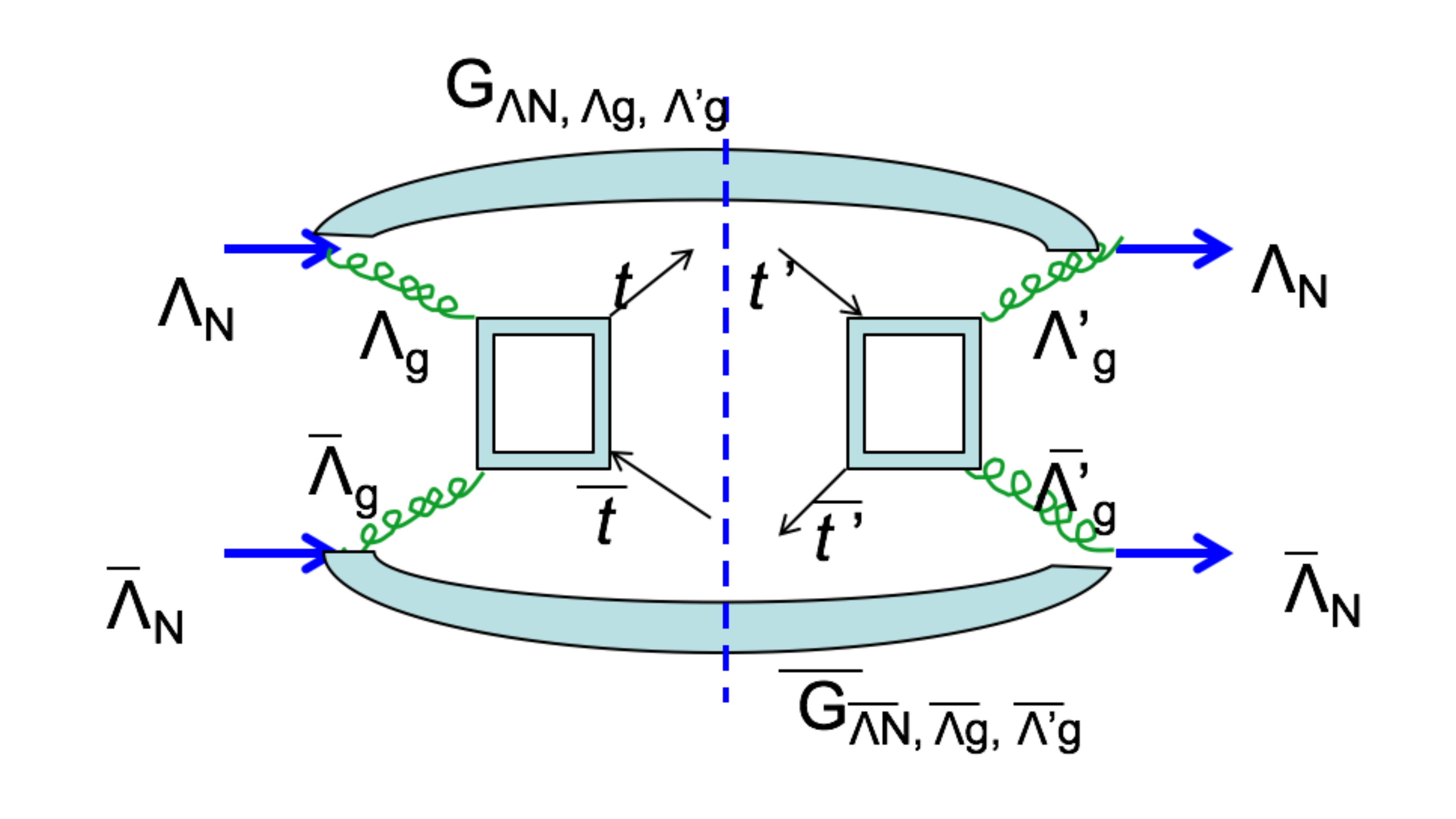}
\caption{Predicted top polarization from p+p collisions at 3.5 TeV vs. $p_T$ at 3 values of $x_F$, based on perturbative QCD model of  Dharmaratna and Goldstein Ref.~\cite{DharmaGoldst}. (left) Illustration of cross section for definite helicities in p+p $\rightarrow$ t+$\bar{\rm t}$ +X. (right)}
\label{t-pol}
\end{figure*}

What is known about single polarization of the top or antitop in the pair production? 
Recent determinations of top single spin asymmetry (SSA) at the LHC are small - from ATLAS $A_p=-0.035\pm0.040$ and from CMS $A_p=0.005\pm0.01$~\cite{ATLAS}.
An explanation based on one loop QCD calculations with an ansatz for ``recombination''~\cite{DharmaGoldst}, predicts
peaks close to $-0.05$ vs. $p_T$, over a range of $x_F$ as shown in Fig.~\ref{t-pol}.
This prediction is within the small values and their uncertainties determined by both CMS and ATLAS.

The spin correlations for the top-antitop pairs produced by unpolarized p+$\bar{\rm p}$ or p+p collisions can be calculated precisely at tree level QCD for the quarks or gluons and folded into the relevant parton distribution functions. At this point there are enough top pairs in the data from ATLAS and CMS to begin to use the spin correlations as probes of the production mechanism.  It has become clear that heavy hadron production~\cite{BoeBro} or even Higgs boson production 
at very high energies could provide a measure of the {\it polarized gluon} distributions in the protons. It is now possible, and especially interesting to use the top pair spin correlation as a lever to disentangle the gluon polarization distributions.

In the following we present the tree level production mechanisms for ${\rm g}_{X\, {\rm or}\, Y}\, + {\rm g}_{X\, {\rm or}\, Y}\, \rightarrow {\rm t}_{\pm}\, + {\bar {\rm t}}_{\pm}\, + {\rm X}$, where the gluon subscripts are linear polarization directions and the t-quark subscripts are helicities. 
The existence of 
top spin correlations at the LHC has been established. A standard parametrization is to represent the top-antitop cross section asymmetries as
\begin{equation}
\frac{1}{\sigma}  \frac{d^2\sigma}{d\cos\theta_1 d\cos\theta_2}=\frac{1}{4}(1 +B_1\cos\theta_1 +B_2 \cos\theta_2 -C_{helicity} \cos\theta_1  \cos\theta_2 )
\label{LHC-param}
\end{equation}
where the polar angles $\theta_1, \, \theta_2$, for the decay product leptons from the top and antitop, are measured relative to the t-direction in the t+$\bar{\rm t}$ center of momentum. 
The measurements of $C_{helicity}$ by ATLAS and CMS agree with the QCD calculations of Ref.~\cite{Bernreuther}.
In general, however, it is the azimuthal dependences that are important for different polarized initial states. The full angular dependences of the top-antitop spin correlations as they depend on gluon distributions are shown in the following.

\section{Gluon-top pair observables}
The QCD amplitudes for g+g$\rightarrow$ t+ $\bar{\rm t}$ are well known at tree level~\cite{DharmaGoldst,parke}. 
Let the amplitudes be $A_{\Lambda_{g1}, \Lambda_{g2} ; \,t, \, {\bar t}}({\hat s},{\hat t})$ with $\Lambda_{g1}, \Lambda_{g2}$ the gluon helicities,  $t\, , {\bar t}$ the top and antitop helicities, and ${\hat s},{\hat t}$ the kinematic invariants in an arbitrary frame. Let $g^{(1)}_{\Lambda_{N1}, \Lambda_{X1}, \Lambda_{g1}}(x_1,k_T,\,M_{X1}^2)$ be the amplitude for proton number 1 to emit a gluon no.1 with longitudinal momentum fraction $x_1$ and transverse momentum $k_T$ with respect to the proton 3-momentum in the collider frame (p + p CM), along with an unspecified residual $X1$ of mass $M_{X1}$, with corresponding $g^{(2)}$ for the other proton. The overall amplitude for N$_1$ + N$_2$ $\rightarrow$ t+ X$_1$ +$\bar{\rm t}$ + X$_2$ is then
\begin{equation}
%\sum \int_{X_1, \,X_2} 
g^{(1)}_{\Lambda_{N1}, \Lambda_{X1}, \Lambda_{g1}}g^{(2)}_{\Lambda_{N2}, \Lambda_{X2}, \Lambda_{g2}}  A_{\Lambda_{g1}, \Lambda_{g2} ; t, {\bar t}} \, .
\label{ttbar_amp}
\end{equation}
To construct differential cross sections for unpolarized colliding protons, this amplitude must be combined with its conjugate and summed and integrated over unobserved quantities as in 
Fig.~\ref{t-pol}. The terms can be rearranged to correspond to gluon distributions and hard scattering amplitude products 
\begin{equation}
G^{(1)}_{\Lambda_{N1},\Lambda_{g1},\Lambda^\prime_{g1}}=\sum_{\Lambda_{X1}} \int_ {X_1} g^{(1)*}_{\Lambda_{N1}, \Lambda_{X1}, \Lambda^\prime_{g1}} g^{(1)}_{\Lambda_{N1}, \Lambda_{X1}, \Lambda_{g1}}  \, .
\label{G1}
\end{equation}

Then the multiple sum can be written more compactly as a double density matrix,
\begin{equation}
\rho_{t^\prime,{\bar t}^\prime; t, {\bar t}}=  \sum_{\Lambda_{g1},\Lambda_{g2},\Lambda^\prime_{g1},\Lambda^\prime_{g2}} \sum_{\Lambda_{N2},\Lambda_{N1}} G^{(2)}_{\Lambda_{N2},\Lambda_{g2},\Lambda^\prime_{g2}}G^{(1)}_{\Lambda_{N1},\Lambda_{g1},\Lambda^\prime_{g1}} A^*_{\Lambda^\prime_{g1}, \Lambda^\prime_{g2} ; t^\prime, {\bar t}^\prime}A_{\Lambda_{g1}, \Lambda_{g2} ; t, {\bar t}} \, .
\label{G1G2}
\end{equation}
 
 When the summations over the various unmeasured helicities are carried out and  parity relations are used, four distinct terms arise, each a variation of the form 
 \begin{eqnarray}
\rho^{LP,LP}_{t^\prime,{\bar t}^\prime; t, {\bar t}}&=&\left[A^*_{LL,t^\prime,{\bar t} }A_{RR,t,{\bar t}} + A^*_{LR,t^\prime,{\bar t} }A_{RL,t,{\bar t}} 
+A^*_{RL,t^\prime,{\bar t} }A_{LR,t,{\bar t}} + A^*_{RR,t^\prime,{\bar t}}A_{LL,t,{\bar t}} \right] \, ,
\end{eqnarray}
so that 
\begin{eqnarray}
\rho_{t^\prime,{\bar t}^\prime; t, {\bar t}}=\sum_{\Lambda_{N1},\Lambda_{N2}}&& \{G^{(2)}_{\Lambda_{N2} ,\, UP} \,\rho^{UP,UP}_{t^\prime,{\bar t}^\prime; t, {\bar t}}\,G^{(1)}_{\Lambda_{N1} ,\, UP} +G^{(2)}_{\Lambda_{N2} ,\, UP} \,\rho^{UP,LP}_{t^\prime,{\bar t}^\prime; t, {\bar t}}\,G^{(1)}_{\Lambda_{N1} ,\, LP} \, + \nonumber \\
& +& G^{(2)}_{\Lambda_{N2} ,\, LP} \,\rho^{LP,UP}_{t^\prime,{\bar t}^\prime; t, {\bar t}}\,G^{(1)}_{\Lambda_{N1} ,\, UP} + G^{(2)}_{\Lambda_{N2} ,\, LP} \,\rho^{LP,LP}_{t^\prime,{\bar t}^\prime; t, {\bar t}}\,G^{(1)}_{\Lambda_{N1} ,\, LP}\} \, .
\end{eqnarray}  

The subscripts $R,L$ correspond to gluon helicities $\pm 1$. Because of parity relations 
the combination of gluon distributions that appear in the summation is limited. The two independent combinations correspond to linear polarization states:
\begin{eqnarray}
G^{(1)}_{\Lambda_{N1},R,R} + G^{(1)}_{\Lambda_{N1},L,L}&=& G^{(1)}_{\Lambda_{N1},XX} + G^{(1)}_{\Lambda_{N1},YY} = G^{(1)}_{\Lambda_{N1} ,\, UP} \, ,
\label{UP} \\
G^{(1)}_{\Lambda_{N1},R,L} + G^{(1)}_{\Lambda_{N1},L,R}&=&G^{(1)}_{\Lambda_{N1},YY} - G^{(1)}_{\Lambda_{N1},XX} = G^{(1)}_{\Lambda_{N1} ,\, LP} \, .
\label{LP}
\end{eqnarray} 
The $UP$ and $LP$ subscripts on the right are for unpolarized and linearly polarized gluons. The ${\hat X}\, \& \,{\hat Y}$ directions are transverse to the gluon 3-momentum ${\vec k}_1$, with ${\hat X}$ in the g$_1$+g$_2$ $\rightarrow$ t+$\bar{\rm t}$ scattering plane. For gluon number 2, the 3-momentum ${\vec k}_2$
is neither parallel nor anti-parallel to ${\vec k}_1$, in general, but the $X-Z$-planes coincide. So the ${\hat X}$ direction for g$_2$ differs from g$_1$, but the ${\hat Y}$ directions coincide. 
 
The tree level hard scattering amplitudes $A_{\Lambda_{g1}, \Lambda_{g2} ; t, {\bar t}}$ can be evaluated in the CM frame in terms of the variables ${\hat s}, \, \theta, \, \beta$ and the color factors for the $(8)\otimes(8)\rightarrow (3)\otimes ({\bar 3})$. 
The azimuthal dependence enters these density matrix elements.
The double density matrix elements will involve  azimuthal dependences $I, e^{\pm 2 i \phi}, e^{\pm 4 i \phi}$. 
\begin{figure}
\begin{center}
\includegraphics[width=14cm]{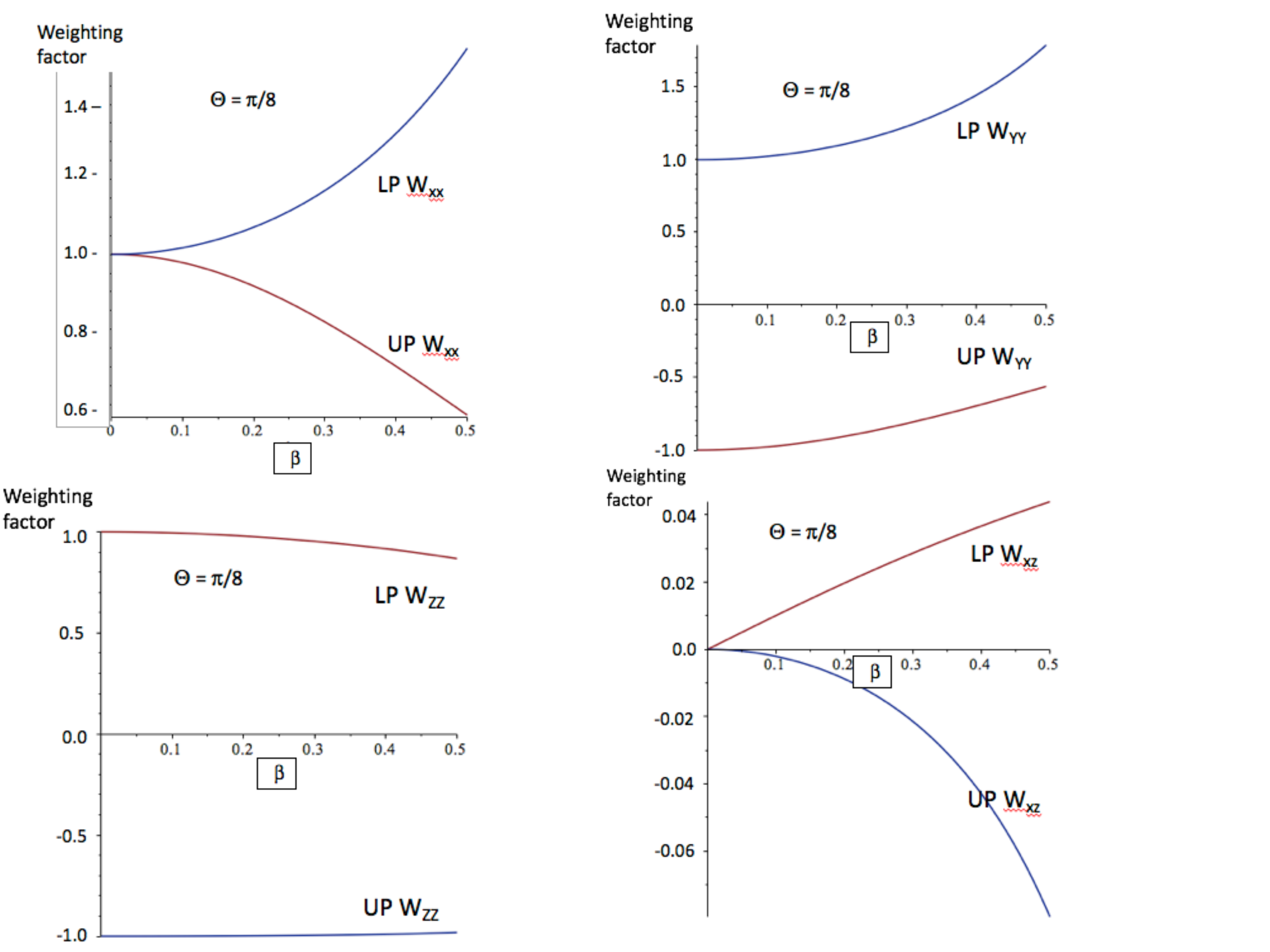}
\end{center}
\caption{Weighting for Cartesian components of $\hat{p}(\mu^+), \; \hat{p}(\mu^-)$, plotted for varying $\beta$, the magnitude of relativistic velocity of the top in the t+$\bar{\rm t}$ Center-of-Mass frame. Each is plotted for unpolarized and transverse-linear polarized gluon distributions. Each lepton momentum is evaluated in the corresponding top or anti-top rest frame, with directions defined by CM.}
\label{weights}
\end{figure}
%%%%%%%%%% continue with azimuthal  
With these amplitudes, and their relations to the gluon helicities or transversities, the top-antitop decays provide markers of polarizations. 

\section{Top decay distributions}

The semi-leptonic decays of the top quark afford the best opportunity for
polarization analysis~\cite{DalGol1}. The opposite-sign leptons
usually have very high transverse momenta and are accompanied by b-quark
jets. So the double correlation of top spins is manifested in the
joint decay distributions into leptons and b-jets. Other decay channels are presented elsewhere.

As shown in Ref.~\cite{DalGol1}, the amplitude 
%$B_{\lambda_b,t}$ 
for a polarized top quark at rest to decay into a measured b-quark and antilepton (or u-quark) along with an unobserved neutrino (or  ignoring the $\bar{d}$ jet) is completely determined in the Standard Model. Since the neutrino is not observed, its 3-momentum is fixed to  lie on an ellipse in a lepton-b-quark coordinate system.  

The quark or gluon spin correlations are transmitted to the decay
products. 
%shown in Fig.~\ref{t-decay}. 
The correlations between the lepton directions and the parent
top spin (in the top rest frame) produce correlations between the lepton
directions, which has been expressed as a weighting factor~\cite{spin96} in the light quark-antiquark annihilation mechanism. 

The gluon fusion mechanism for the weighting factor for unpolarized gluons, is summed over gluon helicities. This gives rise to a fourth order angular distribution: 
%due to the combination of two spin 1 gluons. 
%\bf{GLUON GLUON FUSION}
\begin{eqnarray}
W(\theta,p,p_{\bar l},p_l) & = & \frac{1}{4} - 
\frac{1}{4} 
\left\{ \left[ (1-\beta^2)^2+\sin^4\theta) \right] (\hat{p}_{\bar l})_x(\hat{p}_l)_{\bar{x}} \, + \right. \nonumber \\
&& \left. \hspace{0.5in}  + [-(1-\beta^2)^2 -(1-2\beta^2) \sin^4\theta ] (\hat{p}_{\bar l})_y(\hat{p}_l)_{\bar{y}} \, +
 \right. \nonumber \\
 && \left.  \hspace{0.5in} +[(1-\beta^4) -2 \beta^2 \sin^2\theta +\sin^4\theta ] (\hat{p}_{\bar l})_z(\hat{p}_l)_{\bar{z}} \, + \right. \nonumber \\
&& \left. \hspace{0.7in} +  2(\beta/\gamma) \sin^3\theta \cos\theta   [(\hat{p}_{\bar l})_x(\hat{p}_l)_{\bar{z}} 
- (\hat{p}_{\bar l})_z(\hat{p}_l)_{\bar{x}} ] \right\}  \nonumber \\
&& \hspace{0.7in}  / 4\left[ (1-\beta^4) +2\beta^2 \sin^2\theta+(1-2\beta^2) \sin^4\theta \right] .
\end{eqnarray}
where $m$ is the top quark mass, $\theta$ is the top quark production
angle in the quark-antiquark or ${\bar t} \,  t$ CM frame, $p$ is the light quark or gluon CM momentum, $\beta$ is the magnitude of the relativistic velocity of the top or antitop quark in the CM, 
$\hat{p}_{\bar{l}}$ is the $l^+$ momentum direction in the top rest frame
and $\hat{p}_l$ is the corresponding $l^-$ direction in the antitop rest
frame.

We can now separate the dilepton angular distributions into different components for the four different combinations of gluon distributions.
%, shown in Table~\ref{dtable}. 
For the (LP, LP) case, 
which measures the linearly polarized gluon pair.
\begin{eqnarray}
W^{(LP,\, LP)}(\theta,p,p_{\bar l},p_l) & = &- \frac{1}{4}+\frac{1}{4} \left\{ [(1-\beta^4)+\beta^2 \sin^2\theta(-2+(2-\beta^2)\sin^2\theta)] (\hat{p}_{\bar l})_x(\hat{p}_l)_{\bar{x}} \, + \right. \nonumber \\
&& \left. \hspace{0.5in}  +  [(1-\beta^4)+\beta^2 \sin^2\theta(2-\beta^2\sin^2\theta)] (\hat{p}_{\bar l})_y(\hat{p}_l)_{\bar{y}} \, +
 \right. \nonumber \\
 && \left.  \hspace{0.5in} +[-(1-\beta^2)^2 + \beta^2 (2-\beta^2)\sin^4\theta ] (\hat{p}_{\bar l})_z(\hat{p}_l)_{\bar{z}} \, + \right. \nonumber \\
&& \left. \hspace{0.7in} -4(\beta^2 / \gamma) \sin^3\theta \cos\theta   [(\hat{p}_{\bar l})_x(\hat{p}_l)_{\bar{z}} 
- (\hat{p}_{\bar l})_z(\hat{p}_l)_{\bar{x}} ] \right\}  \nonumber \\
&& \hspace{0.7in}  / \left[ (1-\beta^2)^2 +\beta^4 \sin^4\theta \right] .
\end{eqnarray}
In Figure \ref{weights} we show the directional correlation distributions for an unpolarized gluon distribution and a linear-transverse polarized gluon distribution. We have not included any particular values of gluon distribution functions. For that, we would convolute our spectator model distributions with the weights. The distributions are rich in dependences on the energies and angles for the t + $\bar{\rm t}$ pair and the dilepton momenta. The $W_{i\, j}$ is the $\theta$ and $\beta$ dependent factor multiplying the Cartesian components of $\hat{p}(\mu^+)_i \; \hat{p}(\mu^-)_j$, plotted for varying $\beta$, the magnitude of relativistic velocity of the top in the t + $\bar{\rm t}$ Center-of-Mass frame. The lepton momenta are determined in their top or antitop rest frames. Coordinates in the t + $\bar{\rm t}$ CM are determined as follows. The t + $\bar{\rm t}$ pair have momentum $\vec{p}_{t+\bar{t}}$ in the p+p collider CM. In the t + $\bar{\rm t}$ CM the pair of gluons (or quark-antiquark) has zero 3-momentum, so the orientation of one gluon relative to the top in the t + $\bar{\rm t}$ CM is the angle $\theta$ boosted from the p+p CM. 

For illustration we chose the polar angle of the t + $\bar{\rm t}$ CM to be $\theta=\pi/8$ and varied $\beta$. The resulting weighting factors are remarkable for the clear distinction between unpolarized and polarized gluons.

%\section{Gluon Distributions - GPDs and TMDs}
\section{Gluon TMDs}

Gluon TMDs and GPDs are more complicated at leading twist than their valence quark counterparts. For several of the gluon analogs of the quark TMDs at small x, like the Sivers function, there are two kinds of gluon distributions. Which of the two is probed depends on the process. These distinctions have been developed over time, beginning with the realization that there are two ways to insert gauge links into the hadronic matrix elements of the gluon-gluon correlator~\cite{KKT}:

\be
\Gamma^{\mu \nu [\mathcal{U},\mathcal{U}^\prime]}(x,{\bf k}_T ) \equiv 
\int 
%\int 
\frac{d(\xi\cdot P)d^2\xi_T}
{(P\cdot n)^2(2\pi)^3} 
e^{i(xP+k_T)\cdot \xi} \langle P\mid {\rm Tr}_c [ F^{+\nu}(0) \mathcal{U}_{[0,\xi]} F^{+\mu} (\xi) \mathcal{U}^\prime_{[\xi, 0]} ] \mid P \rangle \mid_{\xi \cdot n = 0} 
\ee
%\end{equation}
%\nonumber  \\

%\mathcal{
\be
 {\rm with} \hspace{0.2in} 
\mathcal{U}_{[0,\xi]} = \mathcal{P} \rm{exp} 
 \left( -ig \int_{\mathcal{C}[0,\xi]} ds_\mu A^\mu (s) \right), 
%\hspace{0.2in} {\rm with} \; \xi = [0^+, \xi^-, {\bf \xi}_T ].
%\end{eqnarray}
\ee
for a contour $\mathcal{C}[0,\xi]$ (wherein the 4-vector notation $  \; \xi = [0^+, \xi^-, {\bf \xi}_T ]$). The path determined gauge links can point to $\xi^- \rightarrow \pm \infty^- $.

At small {\it x}, pairing these two path definitions corresponds to Weiz{\"a}cker-Williams $[+ \, +]$ and dipole $[+ \, -]$  distributions, respectively. For Weiz{\"a}cker-Williams (WW), $\gamma^*$ + g $\rightarrow \gamma^\prime$ + g$^\prime$, the incoming $\gamma^*$ can be replaced by equivalent on-shell gammas for which there will be no initial state interactions. The dipole approximation (DP) has the $\gamma^* \rightarrow$ q +$\bar {\rm q}$ Fock states and the quark or antiquark interacts with the gluon, hence an initial state interaction. These two are distinguished by the color structure, the f-type or d-type coupling, at the tree level. Because the top pair production has to occur at moderate values of x at the LHC, only the WW case is of relevance.

The double helicity flip {\bf does not mix} with quark distributions, which makes gluon transversity unique and useful. In the definition of transversity~\cite{GG-MJM} for on-shell gluons or photons, wherein there are no helicity 0 states, the transversity states are:
\begin{eqnarray}
\mid +1)_{trans} &=& \{ \mid +1 \rangle + \mid -1 \rangle \} /2 = \mid -1 )_{trans} \, , \nonumber \\
\mid 0)_{trans} &=& \{ \mid +1 \rangle - \mid -1 \rangle \} /\sqrt{2} \, , \nonumber \\
{\rm helicity} \quad \mid \pm 1 \rangle &=& \{ \mp \hat{x} - i\hat{y} \}/\sqrt{2} \, .\nonumber 
%\hat{x} &=& -\mid 0 )_{trans} = P_{parallel} \nonumber \\
%\hat{y} &=& i \sqrt{2} \mid +1 )_{trans} = P_{normal}
\end{eqnarray}
where the two-body scattering plane is the X-Z plane, with $\hat{y}$ along the normal to the scattering plane.

In a forthcoming publication~\cite{GLgluons} we will present our explicit model for the gluon GTMDs, generalizing from the Regge-diquark spectator model, the ``flexible model". We will address some questions that are unique to gluon distributions: how are the {\it t} and skewness $\xi$ dependences normalized? How is the small {\it x} behavior accounted for? What is the connection to the Pomeron? 

In hadronic collision processes, the gluon distributions are folded into the more probable initial and final state interactions. Nevertheless, we will see that at the LHC, the production of top pairs can enhance the 
ability to separate out a form of polarized gluon contributions.

\section{Acknowledgements}
Work on gluon distributions was done with the collaboration of S. Liuti, J.O. Gonzalez Hernandez and J. Poage. Some of the content was presented in J. Poage's Tufts University Ph.D. dissertation (2017).
I am grateful to the organizers of SPIN2018 for a productive meeting. 
%Work of S.L. supported by U.S. D.O.E. grant DE-SC0016286.


\begin{thebibliography}{99}

\bibitem{GGL} G.~R.~Goldstein, J.~O.~Hernandez and S.~Liuti,
  %``Flexible Parametrization of Generalized Parton Distributions from Deeply Virtual Compton Scattering Observables,''
  Phys.\ Rev.\ D {\bf 84} (2011) 034007; {\it ibid} Phys. Rev. D {\bf 91} (2015) 114013; 
{\it ibid}, J. Phys. G: Nucl. Part. Phys. {\bf 39} (2012) 115001.

\bibitem{GLgluons} G.R. Goldstein, S. Liuti, J.O.~Gonzalez Hernandez, J. Poage
{\it work in progress}; G.R. Goldstein and S. Liuti, ``QCD Evolution 2014", IJMP: Conf. 37(2015) 1560038. 

\bibitem{D0} S. Abachi, {\it et al.}, Phys. Rev. Lett. {\bf 74} (1995) 2422.

\bibitem{CDF} F. Abe, {\it et al.}, Phys. Rev. Lett. {\bf 74} (1995) 2626.

\bibitem{spin96} Gary R. Goldstein,``{\it Spin Correlations in Top Quark
     Production and the Top Quark Mass \/}''{\it in\/} Proceedings
     of the 12th International Symposium on High Energy Spin Physics,
     Amsterdam, edited by
     C.W. deJager {\it et al.}, World Scientific, Singapore, 1997, p. 328.

\bibitem{ATLAS} (ATLAS) Phys. Rev. D{\bf 93} (2016) 012002; {\it ibid}, Phys. Rev. Lett. {\bf 114} (2015) 142001;
(CMS) S. Chatrchyan, {\it et al.} 
Phys. Rev. Lett. {\bf 107} (2011) 021802; {\it ibid}, Phys. Rev. Lett. {\bf 112} (2014) 182001.

\bibitem{DharmaGoldst} W.G.D. Dharmaratna and G.R. Goldstein, Phys. Rev.{\bf D41} (1990) 1731; {\it ibid}, Phys.Rev. {\bf D53} (1996) 1073; 
%\bibitem{GGLambda_c} 
G.R. Goldstein, 
  Polarization of inclusively produced $\Lambda_c$ in a QCD based hybrid model, 
  in {\it Proceedings  $\Lambda_c$ Workshop},
                (Fermilab (1999)) and arXiv:hep-ph/9907573.

\bibitem{BoeBro} D. Boer, {\it et al.}, Phys. Rev. Lett. {\bf 106} (2011) 132001;
%\bibitem{Boer}
C. Pisano, {\it et al.},
{\it JHEP}{\bf 10} (2013) 024.


\bibitem{Bernreuther} W. Bernreuther, et al., Phys. Rev. Lett. {\bf 87} (2001) 242002.


\bibitem{parke} %S. Parke and Y. Shadmi, Phys. Lett. {\bf B387}, 199
 % (1996); G. Mahlon and S. Parke, Phys. Rev. {\bf D53}, 4886 (1996); 
 % G. Mahlon and S. Parke, Phys. Lett. {\bf B411}, 173 (1997); 
 G. Mahlon and S. Parke, Phys. Rev.{\bf D81} (2010) 074024, and references therein.

\bibitem{DalGol1} R. H. Dalitz and G. R. Goldstein, Phys. Lett. B287 (1992) 225; {\it ibid}
%\bibitem{DalGol2} R.H. Dalitz and G.R. Goldstein, 
Phys. Rev. D45 (1992) 1531.

\bibitem{KKT} D. Kharzeev, Y. V. Kovchegov and K. Tuchin, Phys. Rev. D 68 (2003) 094013.

%\bibitem{Dominguez_etal} F. Dominguez, C. Marquet, B. W. Xiao and F. Yuan, Phys. Rev. D 83 (2011) 105005.

%\bibitem{Hood-Ji} P. Hoodbhoy and X.-D. Ji, Phys. Rev. {\bf D 58}, 054006 (1998).

%\bibitem{Burkardt} M. Burkardt, IJMPA{\bf 18}, 173 (2003).

\bibitem{GG-MJM} G.R. Goldstein and M. J. Moravcsik, 
%Ann. Phys. {\bf 195}, 213 (1989).     
 {\it Ann. Phys.} (N.Y.) {\bf 98} (1976) 128;
  %{\bf 142}, 219 (1982); 
  {\bf 195} (1989) 213. 

\end{thebibliography}
\end{document}